\documentclass[10pt]{article}
\usepackage{amsmath}
\usepackage{amssymb}
\usepackage{graphicx}

\def\aa{A\&A }

\def\aj{AJ }

\def\mnras{MNRAS }

\begin{document}

\setcounter{figure}{0}
\setcounter{table}{0}
\setcounter{footnote}{0}
\setcounter{equation}{0}

\vspace*{0.5cm}

\noindent {\Large TOWARD REINFORCING THE LINK BETWEEN GAIA AND ICRF FRAMES}

\vspace*{0.7cm}
\noindent\hspace*{1.5cm} Z. MALKIN \\
\noindent\hspace*{1.5cm} Pulkovo Observatory \\
\noindent\hspace*{1.5cm} Pulkovskoe Sh. 65, St. Petersburg 196140, Russia \\
\noindent\hspace*{1.5cm} e-mail: malkin@gaoran.ru \\

\vspace*{0.5cm}
\noindent {\large ABSTRACT.}
The link problem between radio (VLBI/ICRF) and optical ({\it Gaia}/GCRF) celestial reference frames is analyzed.
Both systems should be a realization of the ICRS (International Celestial Reference System) at microarcsecond level of accuracy.
Therefore, the link between the ICRF and GCRF should be obtained with similar accuracy, which is not a trivial
task due to relatively large systematic and random errors in source positions at different frequency bands.
In~this presentation, additional possibilities to improve the GCRF-ICRF link accuracy are discussed.
In~particular, a possibility to increase the number of ICRF and GCRF common objects is considered
using advanced scheduling of the regular IVS sessions such as R1 and R4.
It is shown that inclusion of supplement prospective southern sources in these sessions allows enriching southern ICRF zone
without noticeable loss of accuracy of geodetic results.
Another topic discussed in this presentation is using the correlations between radio source coordinates,
which can impact the orientation angles between two frames at a level of a few tens of$\mu$as.

\vspace*{1cm}
\noindent {\large 1. INTRODUCTION}
\smallskip

During last twenty years, the International Celestial Reference System (ICRS) and Frame (ICRF) are based on VLBI observations of extragalactic
radio sources, which currently reached the accuracy of a few tens of~$\mu$as (Fey et al. 2015).
The next, third ICRF realization, ICRF3 (Jacobs et al. 2014) will be published in 2018 and is expected to be approved by the IAU General Assembly
in August 2018.

On the other hand, the {\it Gaia} astrometric satellite was sufficiently launched in December 2013 and already delivered the first scientific
results of revolutionary quality.
In particular, one of the main results of the {\it Gaia} mission will be the optical celestial reference frame GCRF based on positions and velocities
of more than a billion objects and having the accuracy similar to ICRF.

Ideally, both ICRF and GCRF should be realizations of the ICRS (International Celestial Reference System).
However, determination the link parameters between two frames is not a trivial task due to relatively large systematic
and random errors in source positions at different frequency bands.
The goal of this presentation is to draw attention to some possibilities to reinforce the GCRF--ICRF link that has not been fully explored yet.

Among other objects observed with {\it Gaia}, there are about half a million of extragalactic radio sources.
A few thousands of them can be cross-identified with the radio sources having accurate positions obtained from VLBI observations.
These GCRF and ICRF common objects are used to establish a link between optical and radio CRF realizations.
It can be anticipated that increasing the number of common ICRF--GCRF sources will be important for several purposes.
First, it provide better formal precision of the link parameters between two frames.
It is also very important that both frames are subject of systematic errors.
Therefore, the more common objects are used for comparison, the more detailed investigation of these errors can be performed.
Another principal difference between ICRF and GCRF positions is that they are generally referred to different parts of the same object
observed in optics and radio, see discussion in, e.g., Jacobs (2014), Mignard et al. (2016), Petrov \& Kovalev (2017) and papers cited therein.

Taking into account that {\it Gaia} is expected to provide accurate positions of 400--500 thousand of extragalactic objects, while ICRF
contains about four thousand radio sources, it is the ICRF catalog that should provide more objects to improve the link.

Another topic of this work is related to discussion of possible sources of systematic difference between ICRF and GCRF coming from
some yet unresolved problems of frame comparison procedures. 

This presentation is mostly a summary of several previous studies of the author and colleagues on the subject,
such as Malkin et al. (2013), Malkin (2016), Sokolova \& Malkin (2016).
\clearpage

\noindent {\large 2. INCREASING THE NUMBER OF THE LINK SOURCES}
\smallskip

\begin{figure}[ht]
\begin{center}
\includegraphics[clip,scale=1]{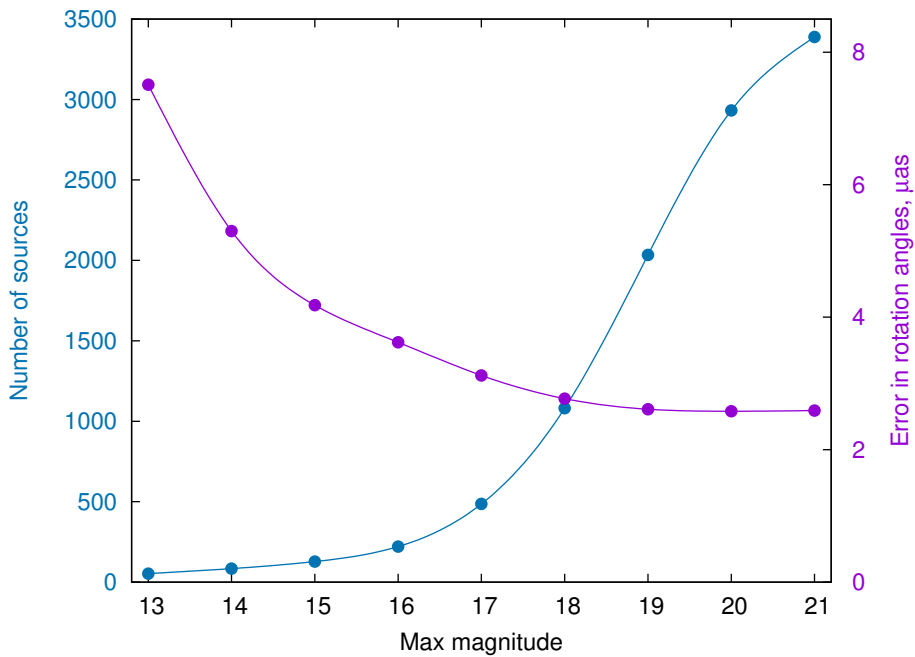}
\caption{Dependence of the number of GCRF--ICRF link sources and the uncertainty in the rotation angles
  between the frames on the upper visual magnitude limit.}
\label{fig:nsou_error_vs_mag}
\end{center}
\end{figure}

One of the objectives of the ICRF3 was to increase the number of radio sources having accurate radio position and sufficiently bright in optics
to be observed with {\it Gaia}.
In the framework of this program about two hundred prospective radio sources of optical magnitude $\le$18 and of good radioastrometric quality
were selected and observed (Bourda et al. 2008, 2010, 2011; Le Bail 2016).
However, the {\it Gaia} magnitude limit is much fainter than $18^m$ accepted in this program.
It was planned before the mission start that {\it Gaia} will observe objects with magnitude $G \le 20$.
Actually, {\it Gaia} counterparts of the astrometric radio sources have $G$ magnitude
up to $\sim$21.5$^m$ for ICRF sources (Mignard et al. 2016).
Figure~\ref{fig:nsou_error_vs_mag} presents dependence of the number of sources and modeled error in the orientation angles on the magnitude limit.
For this simulation, radio source list was taken from the latest official NASA GSFC VLBI group astrometric solution of December
2016\footnote{https://gemini.gsfc.nasa.gov/solutions/2016a\_astro/2016a\_astro.html}.
This GSFC catalog contains positions for 4196 radio sources.
Optical magnitudes for these sources were taken from the OCARS catalog\footnote{http://www.gaoran.ru/english/as/ac\_vlbi/ocars.txt}.

One can see that moving from 18$^m$ to 21$^m$ provides about 3 times more sources and reducing the orientation angles uncertainty by $\sim$10\%.
Analyzing this result, one should bear in mind that the {\it Gaia} position error degrades with increasing the optical magnitude.
Taking into account the general problem of deficiency of the astrometric VLBI resources,
a separate detailed study may be needed to investigate what is the optimal trade-off between the number of the link sources
and their VLBI-derived position error for the accuracy of the link between two frames.
However, a larger number of sources should be always preferable to mitigate the systematic errors of the radio and optical source positions.

\begin{figure}[ht]
\begin{center}
\includegraphics[clip,scale=0.7]{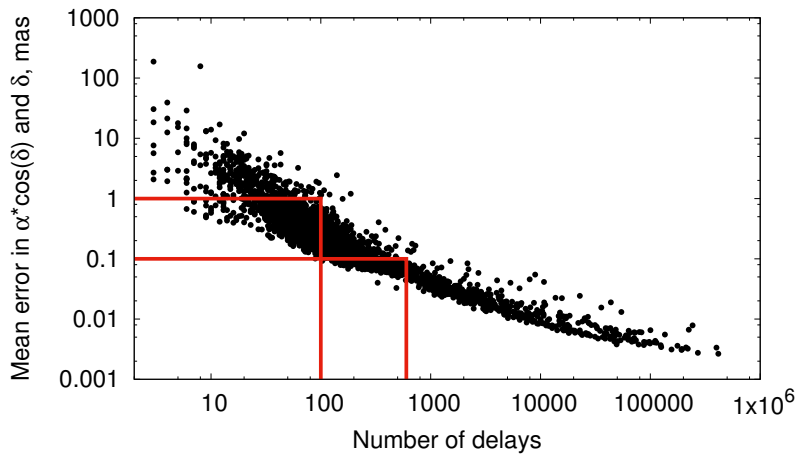}
\caption{Radio source position error vs. the number of observations.}
\label{fig:position_error_vs_nobs}
\end{center}
\end{figure}

Figure~\ref{fig:position_error_vs_nobs} shows the dependence of the radio source position uncertainty on the number of observations for the GSFC catalog.
It is very close to that for ICRF2 catalog (Fey et al. 2015).
The result shows that about 100 observations are needed to reliably provide the source position error better than 1~mas,
and about 600 observations are needed to reliably provide the error better than 0.1~mas.
Let us consider the first variant as a minimal realistic requirement for a reasonable time perspective.
Let us also notice that the International VLBI Service for Geodesy and Astrometry (IVS, Nothnagel et al. 2017)
data archive  currently includes about 3000 sources with $\ge 100$ observations but only 114 of them with $\delta < -40^{\circ}$,
46 with $\delta < -60^{\circ}$, and 12 with $\delta < -75^{\circ}$.

Indeed, VLBI community is striving to increase the number of sources with highly accurate positions
for $\delta < -45^{\circ}$ (VLBA declination southern limit), where the deficiency of the ICRF source distribution is traditionally poor.
In particular, establishing of new telescopes in Australia, New Zealand and South Africa allowed substantial increasing the number
of observations of southern sources (Basu 2016, Plank et al. 2017).
However, yet other possibilities can be considered to get more sources with reliable position in the framework of existing resources.
One of them is inclusion of targeted southern sources in the schedule for regular multi-baseline IVS programs such as weekly R1 and R4
sessions in the first place.

\begin{figure}[ht]
\begin{center}
\includegraphics[clip,scale=0.47]{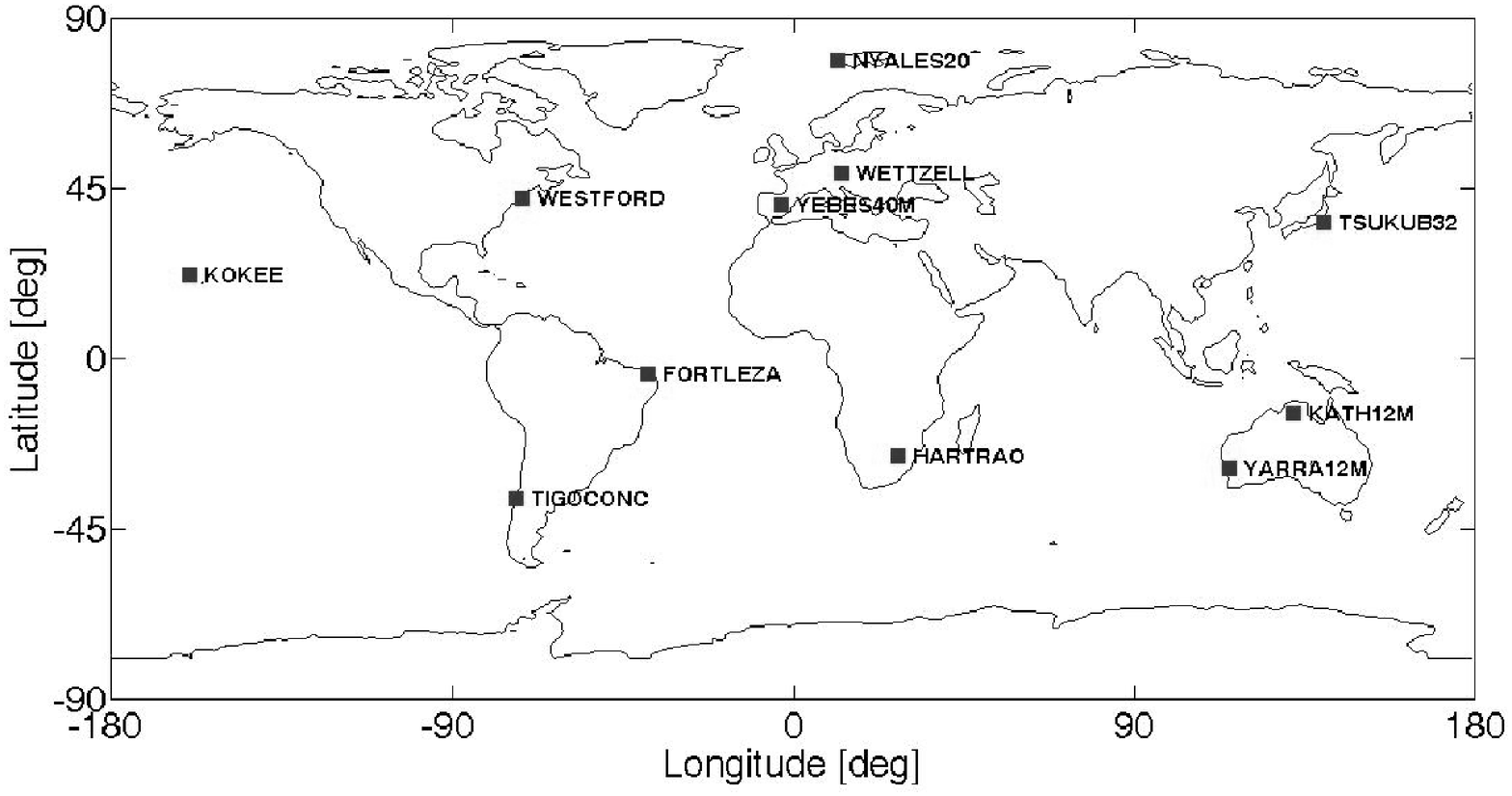}
\caption{11 stations network of IVS R1541 session, 5 out of 11 stations are located in the southern hemisphere.}
\label{fig:r1591_network}
\end{center}
\end{figure}

Simulation of advanced scheduling was done in Malkin et al. (2013).
We started with actual schedule for IVS session R1591 that involved the 11-station network shown in Fig.~\ref{fig:r1591_network}.

\begin{figure}[ht]
\begin{center}
\includegraphics[clip,scale=0.28]{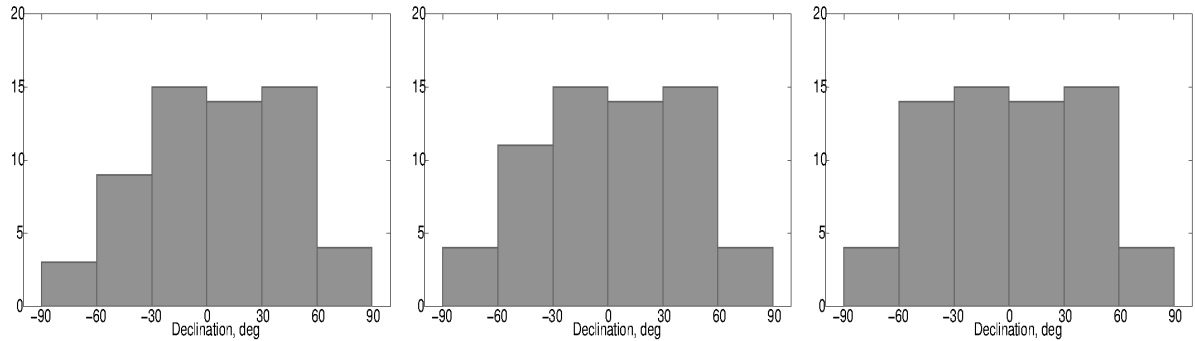}
\caption{Distribution of sources in the original R1541 (R1) schedule (left) and two experimental schedules: R1+ (middle) and R1++ (right).
(Malkin et al. 2013)}
\label{fig:source_dustribution}
\end{center}
\end{figure}

In the original IVS schedule for the R1541 session, 60 sources were observed including 7 southern sources with declination less than $-40^{\circ}$.
For comparisons, the supplementary southern sources are added to the original source list and three experimental schedules were obtained to evaluate
the trade-off between the number of southern sources and the accuracy of geodetic products.
Schedule `R1' was obtained with the original R1591 source list.
Schedule `R1+' includes three more southern sources, and schedule `R1++' includes six more southern sources as compared with the original R1541 schedule.
The three schedules for 24-hour continuous observations were generated with VieVS scheduling package (Sun et al., 2011).
The distribution of the sources in the three schedules is shown in Fig.~\ref{fig:source_dustribution}.

For Monte Carlo simulation, 50 sessions were generated using the same 24-hour schedule but different realizations of noise delays,
each time creating new values for wet zenith delay, clocks and white noise to simulate observations as realistic as possible.
The simulated NGS data files were entered into the software package VieVS (B\"ohm et al. 2012), which computes a classical least squares solution.
The source coordinates were fixed to the ICRF2 positions, and only Earth orientation parameters (EOP) and station positions were estimated.

The standard deviation of the 50 EOP estimates and mean formal uncertainties obtained in our computations are listed in Table~\ref{tab:param-err}.
One can see that we found no overall degradation of the EOP accuracy after the inclusion of supplement southern sources.
Errors in some EOP became even smaller with inclusion of more southern sources, and some EOP showed minor degradation in the accuracy.

\begin{table}
\centering
\caption{Repeatability and standard deviation of EOP for the IVS R1541 and two experimental schedules R1+ and R1++ (Malkin et al. 2013).}
\label{tab:param-err}
\tabcolsep=6pt
\begin{tabular}{lcccc}
\hline
\multicolumn{1}{c}{Parameter} &     &  R1   &  R1+  &  R1++ \\
\hline
Number of scans               &     &  1258 &  1351 &  1375 \\
Number of observations        &     &  3905 &  3813 &  3997 \\
\hline
EOP repeatability             & Xp  & 143.2 & 125.5 & ~98.2 \\
{[$\mu$as, $\mu$s]}           & Yp  & ~98.2 & ~79.1 & ~96.8 \\
                              & UT1 & ~~5.6 & ~~4.6 & ~~5.9 \\
                              & dX  & ~36.2 & ~42.8 & ~39.1 \\
                              & dY  & ~45.0 & ~39.5 & ~37.2 \\
\hline
Mean EOP uncertainty          & Xp  & ~94.8 & ~95.6 & ~93.4 \\
{[$\mu$as, $\mu$s]}           & Yp  & ~77.2 & ~77.3 & ~74.8 \\
                              & UT1 & ~~4.4 & ~~4.6 & ~~4.7 \\
                              & dX  & ~29.8 & ~30.9 & ~29.5 \\
                              & dY  & ~29.1 & ~29.6 & ~28.1 \\
\hline
\end{tabular}
\end{table}

\begin{figure}
\centering
\includegraphics[clip,width=0.8\hsize]{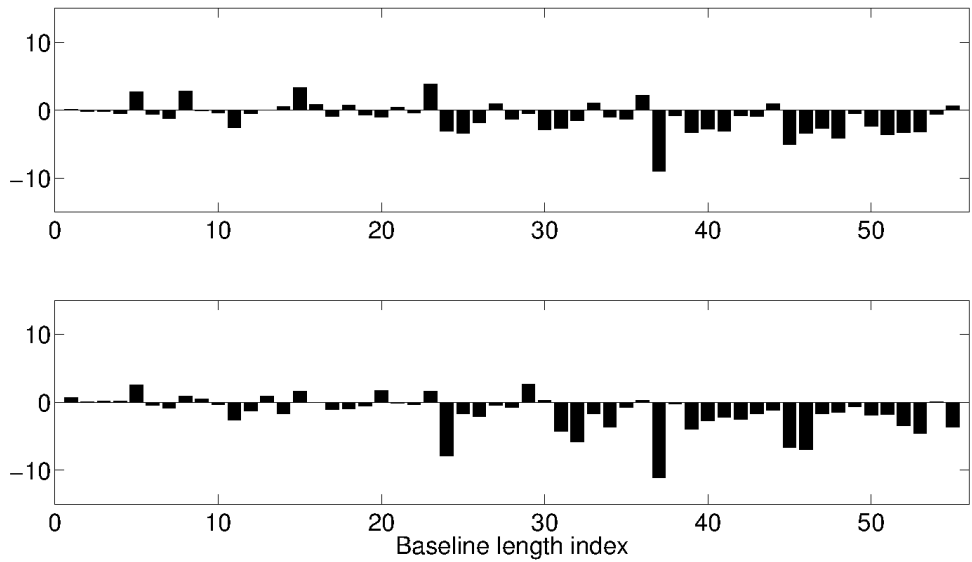}
\caption{Differences in baseline length repeatability [mm] between two schedules: R1+ minus R1 (top) and R1++ minus R1 (bottom).
  The horizontal axis represents the 55 baselines with the shortest one (1575 km) on the left and the longest one (12401 km) on the right.
  (Malkin et al. 2013)}
\label{fig:baseline_length_rep}
\end{figure}

Figure~\ref{fig:baseline_length_rep} shows baseline length repeatability obtained from the simulations.
It was found that for the baselines shorter than $\sim$5,000~km the R1 schedule shows the best result, and R1+ and R1++ schedules shows
worse repeatability, whereas for longer baselines the R1++ schedule is the best, and R1 is the worst.
However, in fact, the results obtained with the three schedules are close to each other.
The mean baseline length repeatability derived from R1, R1+, and R1++ schedules are 13.5 mm, 12.4 mm, and 11.9 mm, respectively.
In other words, increasing of the number of southern sources (cf. R++ and R+ schedules) leads to a small degradation
of baseline length repeatability for short baselines, and small improvement for long baselines.
However, an overall improvement in the baseline length repeatability was found after inclusion more southern sources in the schedule.

Details of this study can be found in Malkin et al. (2013).

Inclusion of targeted southern sources in regular IVS sessions such as R1 and R4 can help to increase substantially the number
of ICRF southern sources with accurate positions.
Suppose, we want to add 100 new sources in the south observed each 100 times (to obtain sub-mas position error) during one year. 
Then we need to make $\sim$200 observations of these sources per week (cf. current $\sim$10K observations in R1+R4 weekly schedule),
which would not significantly influence the normal IVS operations, and might even provide some improvement in obtained geodetic parameters.

\vspace*{1cm}
\noindent {\large 3. USING CORRELATION INFORMATION}
\smallskip

During comparison of the ICRF and GCRF catalogs, a special attention is given to determination of the mutual orientation between two frames
(Mignard et al. 2016).
Jacobs et al. (2010) showed that the result of computation of the orientation angles substantially depend on whether the correlations between
sources coordinates were taken into account.
In their study, the authors compared results of computations of the orientation angles between three own CRF solutions and ICRF2.
They used diagonal or full correlation matrix for their catalogs, and diagonal matrix for ICRF2 (full correlation matrix for ICRF2
is not available).
They found that the orientation angles computed with and without correlations may differ by more than 30~$\mu$as.
Therefore, it is important for the alignment of celestial reference frame (CRF) solutions when a microarcsecond level of accuracy is required,
as is the case for the ICRF--GCRF link.

Sokolova \& Malkin (2016) performed a more detailed investigation of this effect.
They compared results of determination of the orientation angles between different CRF solutions using three methods
of including the correlation information in the computation procedure:
\begin{enumerate}
\item Only position errors are used, which corresponds to one-diagonal covariance matrix.
\item Correlation between right ascension and declination (RA/DE) given in the standard VLBI catalog format is used, which
  corresponds to two-diagonal covariance matrix.
\item Full covariance matrix given in the SINEX solution format is used.
\end{enumerate}

The computations were performed with nine catalogs computed in eight IVS Analysis centres, {\it aus}, {\it bkg}, {\it cgs}, {\it gsf}
(two solutions), {\it opa}, {\it sha}, {\it usn}, and {\it vie}.
Seven catalogs were provided in the standard IERS standard, i.e. included only position errors RA/DE correlations, and two solutions,
{\it gsf} and {\it vie} were kindly provided by their authors in SINEX format, i.e. with full covariance matrix.

Our analysis has shown that using the full covariance matrices leads to substantial change in the orientation parameters between the compared catalogs
({\it gsf} and {\it vie}) of more than 20~$\mu$as, which confirms the result of Jacobs et al. (2010).
On the other hand, using the RA/DE correlations only slightly influences the computed rotational angles.
In some cases, the difference between the first and second variants may reach a few~$\mu$as, but this difference is always smaller than its uncertainty.
Notice that the formal error of the orientation angles in our test computations was typically 2 to 5~$\mu$as.

Another test, similar to the Jacobs et al. (2010) work, was done with {\it gsf} and {\it vie} catalogs.
We compared results of computations of the orientation angles between them and ICRF2.
In this test, all three variants of accounting for the correlations in {\it gsf} and {\it vie} catalogs were tried,
whereas two-diagonal ICRF2 covariance matrix was used in all tests.
In result, we found that difference in orientation angles between the first/second and the third variant can exceed 10~$\mu$as.
Evidently, this test corresponds to the ICRF3--GCRF alignment, provided the ICRF3 catalog is available in SINEX format, and keeping in mind that
the full covariance matrix is not anticipated for the {\it Gaia} catalog.

It is interesting to notice that both our and Jacobs et al. (2010) comparisons of individual catalogs with ICRF2 showed much smaller
(mostly statistically insignificant) orientation angles in the case when the full correlation matrix is used, and the orientation
angles became significantly larger when only one- or two-diagonal matrix is used.
On the other hand, all the individual catalogs used in these tests in both our and Jacobs et al. (2010) works were computed using
the least square adjustment under NNR constraint with respect to ICRF2.
Therefore, we can suggest that during the catalog computation, the least square procedure implicitly uses the full covariance information
contained in the solution.

More details of this study are given in Sokolova \& Malkin (2016).

\vspace*{0.7cm}
\noindent {\large 4. CONCLUSION}
\smallskip

Several conclusions can be drawn from this work.
\begin{enumerate}
\item Including more optically faint radio sources up to 20--21$^m$ in the regular VLBI observing programs provides more accurate and reliable link
of GCRF to ICRF; there is no need to limit the source list to ICRF2.
\item Including prospective southern sources in the regular IVS observing programs like R1 and R4 would allow to substantially improve ICRF
uniformity over the sky.
If this proposal will be realized by the IVS operation centers, up to 100 new sources per year can be added to the ICRF core
without noticeable impact on the geodetic results, such as EOP and baseline length repeatability.
\item The full correlation matrix of the ICRF solution should preferably be used for computation of the GCRF--ICRF orientation parameters.
Therefore, it is strongly advisable that VLBI-based CRF solutions, including ICRF realizations, will be published with the full covariance matrix.
\end{enumerate}

Since the preparation of the ICRF3 is now at the final stage, the proposals discussed in this paper can be hopefully useful for developing
of the next ICRF realization, ICRF4, which can be reasonably planned for 2024 to be compared with the (near)-final version
of the {\it Gaia} catalog presumably at the IAU 2024 General Assembly.

\vspace*{0.7cm}
\noindent {\large 5. REFERENCES}

{

\leftskip=5mm
\parindent=-5mm
\smallskip

Basu, S., de Witt, A., Shabala, S., et al., 2016, ``How Good is the Deep Southern Sky'', In: IVS 2016 General Meeting Proceedings:
"New Horizons with VGOS", Eds. D.~Behrend, K.~D.~Baver, K.~L.~Armstrong, pp. 312--316. 

B\"ohm J., B\"ohm, S., Nilsson, T., et al., 2012, ``The new Vienna VLBI Software VieVS'', In: Proc.IAG Scientific Assembly 2009, IAG Symposia,
Vol. 136,  Eds. S.~Kenyon, M.~Pacino, U.~Marti, pp.~1007--1011, doi: 10.1007/978-3-642-20338-1\_126.

Bourda, G., Charlot, P., Le Campion, J.-F., 2008, ``Astrometric suitability of optically-bright ICRF sources for the alignment with the future
Gaia celestial reference frame'', \aa, 490, pp.~403--408, doi:10.1051/0004-6361:200810667.

Bourda, G., Charlot, P., Porcas, R. W., Garrington, S. T., 2010, ``VLBI observations of optically-bright extragalactic radio sources for
the alignment of the radio frame with the future Gaia frame. I. Source detection'', \aa, 520, id.~A113, doi:10.1051/0004-6361/201014248.

Bourda, G., Collioud, A., Charlot, P., Porcas, R., Garrington, S., 2011, ``VLBI observations of optically-bright extragalactic radio sources
for the alignment of the radio frame with the future Gaia frame. II. Imaging candidate sources'', \aa, 526, id.~A102, doi: 10.1051/0004-6361/201014249.

Fey, A.~L., Gordon, D., Jacobs, C.~S., et al., 2015, ``The Second Realization of the International Celestial Reference Frame by Very Long Baseline
Interferometry'', \aj, 150, id.~58, doi 10.1088/0004-6256/150/2/58.

Jacobs, C.~S., Heflin, M.~B., Lanyi, G.~E., Sovers, O.~J., Steppe, J.~A., 2010, ``Rotational Alignment Altered by Source Position Correlations'',
In: IVS 2010 General Meeting Proc., Eds. D.~Behrend, K.~D.~Baver, pp.~305--309.

Jacobs, C.~S., Arias, F., Boboltz, D., et al., ``ICRF-3: Roadmap to the Next Generation ICRF'',
In: Proc. Journ\'ees 2013 Syst\`emes de R\'ef\'erence Spatio-temporels, Observatoire de Paris, 16--18 Sep, Ed. N.~Capitaine,
Paris Observatory, 2014, pp.~51--56.

Le Bail, K., Gipson, J. M., Gordon, D., et al., 2016, ``IVS observation of ICRF2-Gaia transfer sources, \aj, 151, id.~79, doi: 10.3847/0004-6256/151/3/79.

Malkin, Z., Sun, J., B\"ohm, J., B\"ohm, S., Kr\'asn\'a, H., 2013, ``Searching for an Optimal Strategy to Intensify Observations of the Southern
ICRF sources in the framework of the regular IVS observing programs'', In: Proc. 21st Meeting of the EVGA, Eds. N.~Zubko, M.~Poutanen,
Rep. Finn. Geod. Inst., v.~2013:1, pp.~199--203.

Malkin, Z., 2016, ``Connecting VLBI and Gaia celestial reference frames'', Front. Astron. Space Sci., 3, id.~28. doi: 10.3389/fspas.2016.00028.

Mignard, F., Klioner, S., Lindegren, L., et al., 2016, ``Gaia Data Release 1. Reference frame and optical properties of ICRF sources'', \aa, 595,
id.~A5, doi: 10.1051/0004-6361/201629534 .

Nothnagel, A., Artz, T., Behrend, D., Malkin, Z., 2017, ``International VLBI Service for Geodesy and Astrometry:
Delivering high-quality products and embarking on observations of the next generation'', J. Geodesy,, 91, pp.~711--721, doi: 10.1007/s00190-016-0950-5.

Petrov, L., Kovalev, Y.~Y., 2017, ``On significance of VLBI/Gaia position offsets'', \mnras, 467, pp.~L71--L75.

Plank, L., Lovell, J. E. J., McCallum, J. N., et al., 2017, ``The AUSTRAL VLBI observing program'', J. Geodesy, 91, pp.~803--817,
doi: 10.1007/s00190-016-0949-y.

Sokolova Y., Malkin Z., 2016, ``On the Impact of Correlation Information on the Orientation Parameters Between Celestial Reference Frame Realizations'',
In: IAG 150 Years, Eds. C.~Rizos, P.~Willis, IAG Symposia, 143, pp.~41--44, doi: 10.1007/1345\_2015\_179.

Sun, J., Pany, A.,  Nilsson, T., B\"ohm, J., Schuh, H., 2011, ``Status and future plans for the VieVS scheduling package'',
In: Proc. 20th EVGA Meeting, Bonn, Germany,  Eds. W.~Alef, S.~Bernhart, A.~Nothnagel, pp.~44--48.

}

\end{document}